\documentclass[conference]{IEEEtran}
\IEEEoverridecommandlockouts
\usepackage{cite}
\usepackage{amsmath,amssymb,amsfonts}
\usepackage{algorithmic}
\usepackage{graphicx}
\usepackage{textcomp}
\usepackage{xcolor}
\usepackage{amsmath}
\usepackage{float}
\usepackage{adjustbox}

\usepackage[ruled,vlined,linesnumbered]{algorithm2e}
\def\BibTeX{{\rm B\kern-.05em{\sc i\kern-.025em b}\kern-.08em
    T\kern-.1667em\lower.7ex\hbox{E}\kern-.125emX}}
\begin{document}

\title{Multi-agent Reinforcement Learning-based Network Intrusion Detection System\\

}

\author{\IEEEauthorblockN{Amine Tellache$^{1,2}$, Amdjed Mokhtari$^{1}$, Abdelaziz Amara Korba$^{2}$, Yacine Ghamri-Doudane$^{2}$}
\IEEEauthorblockA{\textit{$^{1}$OODRIVE-Trusted Cloud Solutions, 75010 Paris, France} \\
\textit{$^{2}$L3i Lab, University of La Rochelle, 17000 La Rochelle, France.}\\
emails: \{a.tellache@oodrive.com, a.mokhtari@oodrive.com, abdelaziz.amara\_korba@univ-lr.fr, yacine.ghamri@univ-lr.fr\}
}
}

\maketitle

\begin{abstract}
Intrusion Detection Systems (IDS) play a crucial role in ensuring the security of computer networks. Machine learning has emerged as a popular approach for intrusion detection due to its ability to analyze and detect patterns in large volumes of data. However, current ML-based IDS solutions often struggle to keep pace with the ever-changing nature of attack patterns and the emergence of new attack types. Additionally, these solutions face challenges related to class imbalance, where the number of instances belonging to different classes (normal and intrusions) is significantly imbalanced, which hinders their ability to effectively detect minor classes. In this paper, we propose a novel multi-agent reinforcement learning (RL) architecture, enabling automatic, efficient, and robust network intrusion detection.
To enhance the capabilities of the proposed model, we have improved the DQN algorithm by implementing the weighted mean square loss function and employing cost-sensitive learning techniques. Our solution introduces a resilient architecture designed to accommodate the addition of new attacks and effectively adapt to changes in existing attack patterns. Experimental results realized using CIC-IDS-2017 dataset, demonstrate that our approach can effectively handle the class imbalance problem and provide a fine-grained classification of attacks with a very low false positive rate. In comparison to the current state-of-the-art works, our solution demonstrates a significant superiority in both detection rate and false positive rate.

\end{abstract}

\begin{IEEEkeywords}
Intrusion detection system (IDS), Multi-agent reinforcement learning, Deep Q network (DQN), Class imbalance, CIC-IDS-2017
\end{IEEEkeywords}

\section{Introduction}


Cloud security is a critical aspect of modern-day computing as more organizations rely on cloud services to store, manage, and process their data.
Ensuring cloud security requires a combination of robust security practices. This includes implementing a powerful network intrusion detection system capable of effectively identifying all types of attacks and being adaptable to evolving threats. Signature-based detection and anomaly-based detection are two widely employed approaches for intrusion detection \cite{spadaccino2022intrusion}. Signature-based detection involves searching for specific patterns or signatures of known malicious code or behavior. This method works by comparing incoming data or traffic against a database of pre-existing signatures. 
If a match is found, the system can take action to block or quarantine the malicious code or activity. However, these solutions do not offer automatic or generic detection methods. As a result, they are unable to detect threats that do not match any existing signatures in their databases. 

In order to overcome the limitations of signature-based detection, the anomaly-based detection has appeared. This approach harnesses the power of machine learning (ML) to automatically identify possible intrusions by detecting patterns or behaviors that deviate from the expected patterns. Several machine learning methods have been employed for anomaly-based detection \cite{tama2022systematic, garg_hybrid_2019, kale_hybrid_2022}. However, state-of-the-art benchmark datasets for intrusion detection exhibit class imbalances, where the amount of normal traffic far exceeds that of the attack traffic, mirroring real-world network conditions. Furthermore, within the various attack types, certain attacks occur more frequently than others, exacerbating the issue. Consequently, these imbalances present significant challenges for the IDS in effectively detecting attacks from minor classes, thereby diminishing their overall performance \cite{abdelkhalek2023addressing}. This leads to a high number of false alerts, undermining the IDS effectiveness. 



%

One of the major challenges facing machine learning-based intrusion detection systems is their ability to adapt and evolve with the continuous evolution of attack patterns and the emergence of new attack types. Traditional supervised learning relies on a fixed set of labeled examples for training. However, when the data distribution changes or new types of data are introduced, the performance of the model can deteriorate. Similarly, unsupervised learning aims to discover patterns in a fixed dataset and may struggle to adjust to changes in the initial distribution \cite{ma_aesmote_2021}. Furthermore, most of the existing ML-based IDS \cite{korba2020anomaly, rahal2022antibotv} often require the retraining of the entire model for updates. However, frequent model retraining consumes substantial computing and storage resources, resulting in high costs and time requirements  \cite{jin2023federated}. Unfortunately, only a few studies \cite{jin2023federated} have explored the possibility of incrementally updating intrusion detection systems. In light of the ever-changing landscape of attacks and their continuous evolution, it is crucial for IDS to be purposefully designed to function optimally within dynamic and unpredictable environments.


To neutralize the above-mentioned problems of ML IDS adaptability and class imbalance, we present a novel multi-agent reinforcement learning architecture aimed at enhancing network intrusion detection. Our system harnesses the learning capabilities of RL agents through interactions with the environment, enabling it to effectively address the challenges posed by evolving attack scenarios. RL is specifically tailored to excel in dynamic and uncertain settings, allowing for adaptation to changes in reward and penalty distributions over time. While RL does not explicitly tackle class imbalance, we handle this issue by thoughtfully designing the reward function. Our approach emphasizes the significance of accurately predicting samples from the minority class, thereby encouraging the RL agent to allocate more attention to this class and achieve a better balance.  In summary, the main contributions of this paper are:

\begin{itemize}

\item We propose a modular and adaptable system architecture for our IDS, featuring a flexible and evolvable system architecture. Our design incorporates two levels of Reinforcement Learning: a collection of N independent RL agents, each dedicated to detecting a specific type of attack, and a decision-maker agent responsible for the final classification. This architecture is designed to enable seamless updates in response to newly emerging attacks, eliminating the requirement to retrain the entire system. 




\item We propose a refined version of the Deep Q network algorithm \cite{mnih_playing_2013} specifically tailored to address class imbalance.  Our refined approach incorporates two key enhancements: the integration of the weighted mean square loss function and the implementation of cost-sensitive learning.


\item The experiments conducted on a public dataset demonstrate that our IDS can achieve intrusion detection with an overall accuracy of 99\% and  0.16\% of false positive rate.

\end{itemize}

The remainder of this paper is outlined as follows. Section~\ref{sec:back} presents background on reinforcement learning. Section~\ref{sec:RL} provides an overview of the related work. The proposed multi-agent reinforcement learning-based detection architecture is described in Section~\ref{sec:SOL}. Section~\ref{sec:exp} presents the experimental results and analysis. Finally, Section~\ref{sec:con} concludes the paper and points out future work.


\section{Background} \label{sec:back}

Reinforcement Learning is an interactive learning method focused on decision-making and policy development. 
In RL, an agent learns actions to take based on experiences in order to optimize a quantitative reward over time. The agent operates within an environment and makes decisions based on its current state. In turn, the environment provides the agent with a reward, which can be positive or negative. Through repeated experiences, the agent seeks an optimal decision-making behavior referred to as a strategy or policy that maximizes the cumulative rewards over time \cite{sutton_reinforcement_2018}.

\noindent In the following, we give an overview of the RL methods used in our solution: Q-learning, Deep Q-learning and multi-agent reinforcement learning.
\subsection{Q-learning \& Deep Q-learning}
Q-learning is one of the popular reinforcement learning algorithms used in machine learning. It uses a Q-function to evaluate the quality of an action in a given state. The Q-function is defined as the expected sum of future rewards obtained by taking a particular action in a given state. It can be expressed by \ref{Q-function}: 
\begin{equation}
    Q(s_t, a_t) =\mathbb{E} \; [R_{t+1} + \gamma \max_{a\in A}  Q(s_{t+1}, a)]
    \label{Q-function}
\end{equation}
where $s_t$ is the current state, $a_t$ is the action taken, $s_{t+1}$ is the next state, a is the next action taken, $R_{t+1}$ is the immediate reward obtained from taking action $a_t$ in state $s_t$, $\gamma$ is the discount factor, and $max_{a\in A}  Q(s_{t+1}, a)$ represents the maximum expected reward that can be obtained by taking any action a in state $s_{t+1}$.

The Q-learning algorithm updates the Q-function iteratively using the Bellman equation \ref{Update-function} until it converges to the optimal values for each state-action pair:

\resizebox{\linewidth}{!}{$
    Q(S_t, A_t) \leftarrow  Q(S_t, A_t) + \alpha[R_{t+1} + \gamma \max_{a\in A}  Q(S_{t+1}, a) -  Q(S_t, A_t)]
$}
\label{Update-function}

where $\alpha$ is the learning rate that controls how quickly the Q-function is updated, and the expression in brackets is the TD error, which represents the difference between the expected reward of taking action $a$ in state $s$, and the actual reward obtained plus the estimated future rewards.

The concept of deep Q-learning was first introduced by Volodymyr Mnih, and al \cite{mnih_playing_2013} in 2013. 
It uses a deep neural network, known as the Q-network, to approximate the Q-values for each state-action pair. The deep Q-network takes the state as input and outputs the Q-values for all possible actions, allowing it to handle high-dimensional input spaces. Deep Q-learning leverages a technique called experience replay, where it stores past experiences (state, action, reward, next state) in a replay buffer and randomly samples batches of experiences to train the neural network. This allows the algorithm to break correlations between consecutive samples and improve learning stability. Deep Q-learning with neural networks enables the algorithm to generalize across similar states and handle complex environments.

\subsection{Multi-agent reinforcement learning}
Multi-agent reinforcement learning MARL \cite{busoniu_multi-agent_2010} is an extension of single-agent reinforcement learning.
In multi-agent scenarios, the actions of one agent can have an impact on the rewards and outcomes of other agents, leading to complex interactions and dependencies. The representation of a problem in a multi-agent setting is determined by the nature of interactions between agents, including cooperative, competitive, or mixed, as well as whether agents take actions sequentially or simultaneously. These factors play a crucial role in determining the problem formulation and solution approach within a multi-agent framework \cite{wong2022deep}.


\section{Related Work} \label{sec:RL}

The development of Intrusion Detection Systems (IDS) for cloud environments has garnered considerable attention in recent years due to its crucial role in ensuring the security of these systems. Numerous ML-based IDSs have emerged as potential solutions for effectively detecting attacks in the cloud environment. 
Garg et al. \cite{garg_hybrid_2019} proposed a hybrid algorithm based on the Grey Wolf Optimization (GWO) metaheuristic for multi-objective feature extraction, followed by CNN (Convolutional Neural Network) for anomaly classification. Another hybrid anomaly detection platform \cite{wang_cloud_2022} has been proposed based on SCAE (Stacked Contractive Autoencoder) for feature selection, followed by an SVM model for attack classification. Rahul er al. \cite{kale_hybrid_2022} also proposed a hybrid model, which is divided into several stages of supervised and unsupervised learning. The first stage separates the most obvious anomalies from normal samples using K-means clustering. The remaining "normal" samples are sent to the second stage to identify anomalies using the GANomaly method. A CNN classification model is employed in the final phase for anomaly classification. 
However, in dynamic and evolving environments, the effectiveness and adaptability of methods based on supervised and unsupervised learning can be hindered, particularly when the dataset changes or new patterns are introduced that alter the underlying distribution.



Several variants of reinforcement learning have been proposed for intrusion detection, leveraging its exceptional adaptability, especially in this highly variable field, where attacks continue to evolve and change behavior. Adversarial reinforcement learning has been proposed  to address class imbalance in intrusion detection. The AE-RL model proposed in  \cite{caminero_adversarial_2019} consists of two agents, an environment agent that selects samples for the next training episode, and a classifier agent. Through experience, the environment agent tries to develop a strategy for selecting samples from untreated classes. The classifier agent is responsible for classifying the samples chosen by the first agent. A contradictory configuration is set up between the two agents. Both agents receive contradictory rewards, so that the first agent tries to increase the difficulty of the classifier's prediction by choosing untreated samples. This method has proved effective in dealing with the class imbalance problem. An improved model called AESMOTE, proposed in \cite{ma_aesmote_2021}, builds upon the AE-RL method\cite{caminero_adversarial_2019} by adding over-sampling and under-sampling techniques to further reduce the class imbalance effect. However, adversarial reinforcement learning (ARL) may suffer from convergence problems due to the complex interaction between the agents. It can be difficult to find a balance, which is shown by the reduced precision obtained.


Lopez-Martin et al. \cite{lopez-martin_application_2020} applied different types of DRL algorithms, including Deep Q-Network (DQN), Policy Gradient (PG), and Actor-Critic (AC). The evaluation revealed that the Double Deep Q-Network (DDQN) algorithm outperformed the other algorithms and produced the best results. However, it does not address the problem of class imbalance. Kamalakanta et al. \cite{sethi_deep_2020} combined reinforcement learning with previously trained supervised learning classifiers for intrusion detection. The RL model serves as a validator for predictions made by different classifiers based on predefined thresholds for each classifier. This approach has proved to be effective in achieving the goal of increasing accuracy while maintaining a low false positive rate. However, Adapting this method to newly identified attacks and patterns is difficult due to the necessity of retraining all supervised and reinforcement learning models. Furthermore, the work \cite{sethi_context-aware_2020} proposes a distributed deployment of the solution presented previously in \cite{sethi_deep_2020}. Multiple independent agents will be deployed at the router level based on the network structure and various contexts. A central IDS is implemented to serve as an environment for the reinforcement learning agents. However, this solution does not take into account the computational costs associated with deploying multiple agents, where each agent is a combination of multiple supervised classifiers and reinforcement learning.

In \cite{ramana_ambient_2022}, a new two-level deep reinforcement learning application was proposed for intrusion detection in IoT (Internet of Things) and WSN (Wireless Sensor Networks) environments. The RL-DQN model in the MEN (Mobile Edge networking) layer classifies network traffic as normal or attack, while the DPS (Data processing and storage) layer model subsequently performs a detailed analysis and classifies network traffic into normal category and different types of attacks. Nevertheless, this method does not consider class imbalance issues.



Other models based on multi-agent reinforcement learning systems have also appeared to benefit from the advantages of the MARL architecture compared to single agent. Authors in \cite{sethi_attention_2021} propose an intrusion detection solution based on multi-agent deep reinforcement learning by deploying DQN agents in the different nodes of the distributed network. Two RL levels are implemented, a multi-agent system is deployed at the router level and a final agent for the second level. An improvement has been proposed for the multi-agent DRL by integrating an agent weighting mechanism (attention mechanism). The model learns to associate attention values to each agent in a network according to their performance, the more the agent is efficient, the higher its attention value is. The system includes a central IDS agent that decides whether the traffic is malicious. However, The proposed solution uses a distributed architecture but ignores the computational and network costs by implementing DRL agents at the routers. Authors in \cite{shi_collaborative_2021} propose also an intrusion detection solution based on multi-agent deep reinforcement learning (Major-Minor-RL). The model consists of a major agent and several minor agents. The role of the major agent is to predict whether the traffic is normal or abnormal, while the minor agents are auxiliary to the major agent and help it to correct errors. If the action of the major agent is different from the behavior of most minor agents, the final action will be determined by the minor agents. However, this method does not provide a solution to the class imbalance problem.

Our proposed approach aims to surpass the limitations of existing works. It not only handles the challenges associated with class imbalance issues but also exhibits adaptability towards evolving attack patterns. Furthermore, our IDS seamlessly integrates novel types of attacks.

\section{Proposed Intrusion Detection System} \label{sec:SOL}
In this section, we first present an overview of the proposed multi-agent reinforcement learning model. After that, we explain in detail each element of the training process and we describe the improvements made to the DQN algorithm. 

\subsection{Overview of the proposed appraoch}
\begin{figure}[hbt!]
  \centering
  \includegraphics[width=9cm]{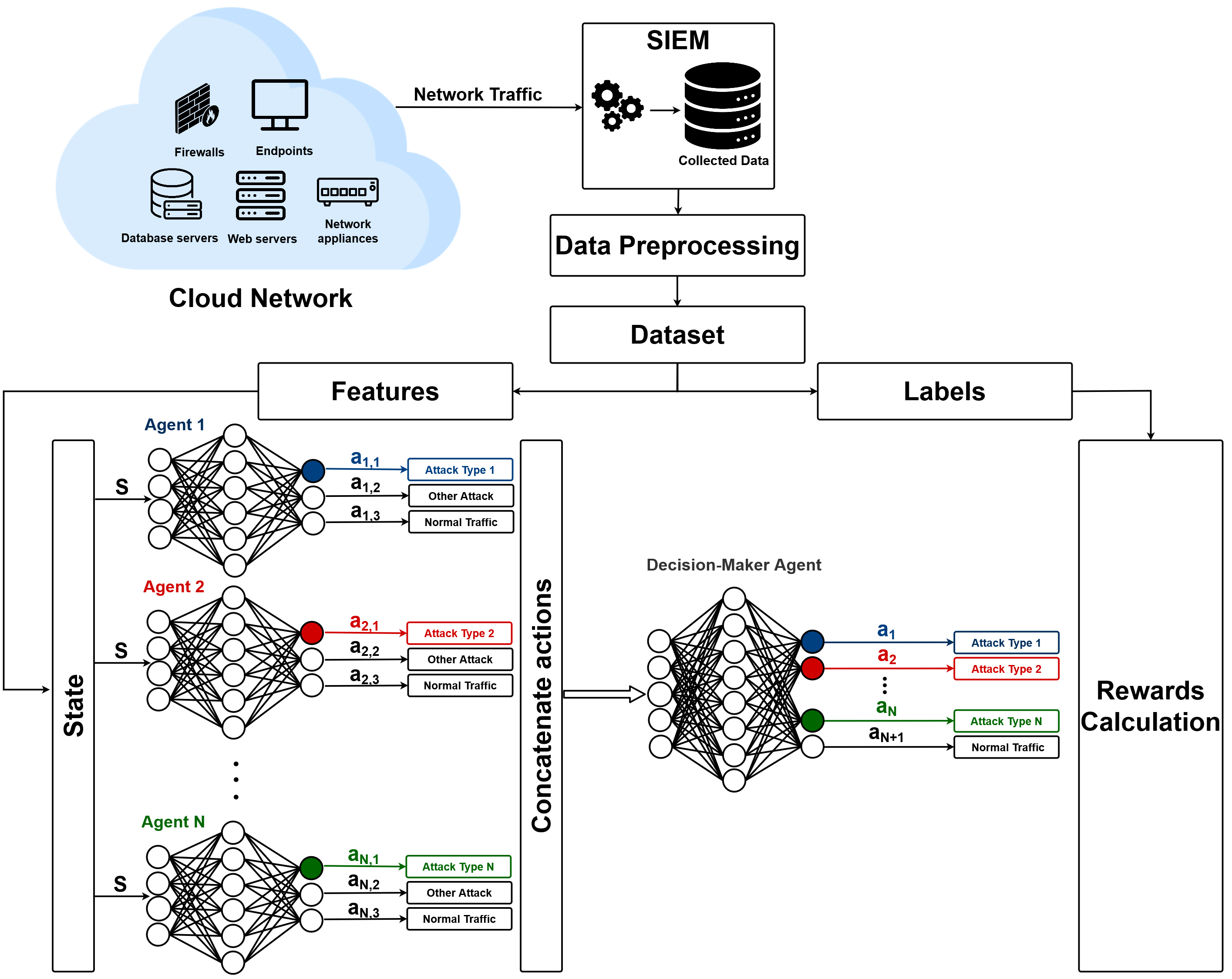}
  \caption{Proposed multi-agent reinforcement learning IDS architecture}
  \label{fig:Proposed-MARL-Architecture}
\end{figure}

We propose a novel approach for intrusion detection utilizing multi-agent reinforcement learning. Our solution, depicted in Figure~\ref{fig:Proposed-MARL-Architecture}, employs a two-level reinforcement learning framework. The first level, known as the detection level, comprises N independent RL agents ($L_1$ agents), with each agent specifically designed to detect a particular type of attack. Each $L_1$ agent is configured with three possible actions: target attack, other attacks, and normal traffic. It is noteworthy that all $L_1$ agents share the same state, which encapsulates all the relevant characteristics of the network traffic.
The second level of reinforcement learning is represented by a single decider agent. The decider agent receives as input the $L_1$ agents outputs (actions of the first level). The decider agent is in charge of giving the final classification of the attack. The set of possible actions for the decider agent includes all the attacks and one action for normal traffic.

Our solution presents a flexible and evolvable architecture, allowing to incorporate new attacks, simply by adding a new $L_1$ agent for the specific attack and re-training the decision agent. Our solution is capable also to handle evolution in existing attack patterns, as it uses reinforcement learning that can adapt to changes in the data distribution. We simply need to retrain the attack agent involved in the evolution and the decision-making agent for a small number of episodes.

Data collection is performed by the SIEM (Security information and event management) \cite{gonzalez-granadillo_security_2021} from the various cloud network sources.
Then, pre-processing is performed on the data by transforming the raw data into exploitable data using cleaning and normalization techniques.
During the "training" process, the dataset is divided into two elements, the feature set and the corresponding labels.
The Feature set represents the state of the $L_1$ agents.
The Labels will be used in the calculation of the rewards by comparing the outputs of the different agents with the current label.

\subsection{Model description}
In Algorithm \ref{alg:Multi-Agent-DQN-Algorithm} we present our multi-agent deep reinforcement learning algorithm for intrusion detection.
We start by initializing both the neural network with random weights and the replay memory with capacity $M$ for all agents including the decider.
The replay memory is used to store the agent's experiences, which are then used to train the neural network.
After that, for each training episode, we train each $L_1$ agent to all samples in the dataset. For each record in the dataset, we begin by initializing the current state $s$ of the system, which represents the set of features associated with the current record. Then, an action will be selected, which consists in giving a classification of the record $s$ following $\epsilon$-greedy exploration strategy. As described in lines 6-7 the agent chooses a random action with a probability $\epsilon$ (exploration) otherwise chooses the optimal action from the neural network. It should be noted that the factor $\epsilon$ is initialized to the value 1 to promote the exploration especially that the agent does not have yet a knowledge of the environment at the beginning of the algorithm, and after each episode this factor $\epsilon$ is decreased until reaching 0 to allow the exploitation of the acquired knowledge. Then, referring to lines 8-9, we calculate the reward by comparing the action $a$ with the label of the current state record $s$ (the reward functions are defined in detail later). Each tuple of experience (state, action, reward) collected will be stored in the replay memory.
And finally a minibatch of experiences from the replay memory will be randomly selected to update the neural network. After training all the $L_1$ agents, and saving for each record all the $L_1$ actions that will constitute the state of the decision agent, we run the Algorithm \ref{alg:Decider-Agent-DQN-Algorithm} to train the decision agent similar to $L_1$ agents. \\
The detailed definition of the state space, action space and reward functions for the level 1 agents and the decision maker agent are given below: \\
\textbf{State space (S):} All  $L_1$ agents receive the same state, which consists of the different dataset features. The state space for the decider agent is composed of the outputs (actions) generated by all $L_1$ agents.
Formally, the state vector for $L_1$ agents : $S_t = (x^{t}_1, x^{t}_2,.., x^{t}_k)$, where \textbf{\textit{k}} is the number of features and \textbf{\textit{t}} indicates the element \textbf{\textit{t}} in the dataset. The outputs of each agent consists of three probability values (Q values) agent attack, other attack and normal traffic. These probabilities values of each agent are then concatenated to obtain the state of the decision maker agent, the state vector is defined as follows: 
$S_t = [Q(S_t, a_{1,1}),Q(S_t, a_{1,2}),Q(S_t, a_{1,3}), Q(S_t, a_{2,1}),\\Q(S_t, a_{2,2}),Q(S_t, a_{2,3}),..., Q(S_t, a_{n,1}),Q(S_t, a_{n,2}),Q(S_t, a_{n,3})]$, where \textbf{n} represent the number of agents. \\
\textbf{Action space (A):}
An action is a decision taken by the agent to classify the network traffic, three possible actions for each level 1 agent, the first action is the agent's class attack, the second action corresponds to identifying all other attacks. Finally, the third action corresponds to classifying the traffic as normal. Formally, the action vector for each agent $A_k = (a_{k,1}, a_{k,2}, a_{k,3})$, where \textbf{\textit{k}} is the agent number.
The set of actions of the decision agent includes one action for each attack and one action for normal traffic, $A = (a_1, a_2,..., a_k, a_{k+1})$, where \textbf{\textit{k}} is the number of attacks and $a_{k+1}$ is normal traffic action. \\
\textbf{Reward (R):}
Rewards are used to guide the learning process of the agent, by providing positive feedback for correct actions that move it closer to the goal and negative feedback is given for incorrect actions, in our case the goal is to provide the right prediction of the attack.
Agents receive independent rewards by comparing the chosen action, which consists in giving a classification of the network traffic, with the current label.    
The reward policy for each $L_1$ agent is given by :

$$
R =
\begin{cases}
k & \text{if  label = agent class \& action = label} \\
-k & \text{if  label = agent class \& action $\ne$ label}  \\
1 & \text{if  label $\ne$ agent class \& action = label}  \\
-1 & \text{if  label $\ne$ agent class \& action $\ne$ label} \\
-k & \text{if  label $\ne$ agent class \& action = agent class } \\
\end{cases}
$$
\[ where \quad k > 1\]

\noindent We used \textbf{cost-sensitive learning} to assign rewards to $L_1$ agents. We created a reward function that gives higher positive and negative feedback values \textbf{\textit{k -k}} to the correct and incorrect prediction for agent class attack, with the aim of creating specialized agents for each attack type.
Each agent is capable of separating its attack class from other classes, so the problem of detecting and classifying attacks can be shared between several agents, with each agent specializing in a particular attack, which leads to better detection and more accurate classification of attacks, despite the huge imbalance between classes.

\noindent The reward policy for the decision-maker agent is given by :

$$
R =
\begin{cases}
1 & \text{if  action = label} \\
-1 & \text{if  action $\ne$ label}  \\
\end{cases}
$$

\noindent The choice of a loss function is crucial in machine learning and optimization tasks. The loss function measures how well a machine learning model performs on a given task by quantifying the discrepancy between the predicted outputs of the model and the true values or labels. \noindent In order to handle the class imbalance and provide a fine-grained classification, we proposed an improved loss function for the DQN algorithm. We have employed a \textbf{weighted mean square loss} function, represented by Equation \ref{Weighted-mean-square-loss}, to accurately quantify the difference between the predicted Q-values and the target Q-values throughout the training procedure.

\begin{equation}
WMSE = \frac{1}{N} \sum_{i=1}^{N} \left( (Q(s_i,a_i) - q\_target(s_i,a_i)) \cdot w_i \right)^2 
\label{Weighted-mean-square-loss}
\end{equation}
where 
\[
q\_target(s_i,a_i) = R_{i} + \gamma \max_{a\in A} Q(s_{i+1}, a)
\]
The discount factor, denoted by $\gamma$, assumes a value close to zero in our scenario. This choice is motivated by the lack of correlation between states, where each step solely involves predicting the class of attack.

\noindent We assign a weight to each experience stored in the replay memory:
\begin{center}
\[
\begin{array}{cccc}
s_0 & a_0 & r_0 & w_0 \\
s_1 & a_1 & r_1 & w_1 \\
\vdots & \vdots & \vdots & \vdots \\
s_N & a_N & r_N & w_N \\
\end{array}
\]
\end{center}

\noindent In the case of $L_1$ agents, we assign greater importance to samples that align with the agent's attack class. This approach aims to enhance the detection capabilities for each specific attack, ultimately facilitating the development of specialized agents tailored to different attack types. As for the decision-maker agent, we prioritize higher weights for minority classes, addressing the issue of class imbalance. This strategy enables effective handling of imbalanced class distributions.

%



\begin{algorithm}[h]
    Initialize the replay memory $\mathcal{D}$ to capacity $\mathcal{M}$ of all $L_1$ agents and Decider agent \;
    Initialize Q-network for all agents with random weights \;
      
    \For{i = 1,2,...,Nb\_Episodes}{        
        \For{each Agent in $L_1$ Agents}{
                \For{each sample $S_k$ in dataset}{
                        Select random action $a^{(k)}$ with probability $\epsilon$ \;
                        Otherwise select $a^{(k)} = argmax_a Q(s^{(k)},a^{(k)};\theta)$ with probability $1 - \epsilon$ \;
                        Store $S\_central\_agent^{(Agent, k)} = \{Q(s^{(k)},a^{(0)}), Q(s^{(k)},a^{(1)}), Q(s^{(k)},a^{(2)})\}$
                        Compare $a^{(k)}$ with the actual label of the sample and calculate $r^{(k)}$ \;
                        Store transition $(s^{(k)},a^{(k)},r^{(k)},s^{(k+1)})$  in $\mathcal{D}$\;
                        Sample random minibatch of transitions $(s^{(k)},a^{(k)},r^{(k)},s^{(k+1)})$ from $\mathcal{D}$ \;
                        Set $y^{(j)} = r^{(k)} + \max\limits_{a^{(j+1)}} Q(s^{(j+1)},a^{(j+1)}; \theta)$\;
                        Perform Stochastic Gradient Descent update on $\theta$ according to the loss function (equation \ref{Weighted-mean-square-loss}) \;
                }
            }
        
        \textbf{Decision\_Maker\_Agent(}$S\_central\_agent$\textbf{)} 
        }
  \caption{Multi-Agent DRL-based IDS}
  \label{alg:Multi-Agent-DQN-Algorithm}
\end{algorithm}

\begin{algorithm}[h]
    Input : actions of the $L_1$ multi-agent-DQN $S\_central\_agent$ \;
    \For{each sample $S_k$ in $S\_central\_agent$}{
        Select random action $a^{(k)}$ with probability $\epsilon$ \;
        Otherwise select $a^{(k)} = argmax_a Q(s^{(k)},a^{(k)};\theta)$ with probability $1 - \epsilon$ \;
        Compare $a^{(k)}$ with the actual label of the sample and calculate $r^{(k)}$ \;
        Store transition $(s^{(k)},a^{(k)},r^{(k)},s^{(k+1)})$  in $\mathcal{D}$ \;
        Sample random minibatch of transitions $(s^{(k)},a^{(k)},r^{(k)},s^{(k+1)})$ from $\mathcal{D}$ \;
        Set $y^{(j)} = r^{(k)} + \max\limits_{a^{(j+1)}} Q(s^{(j+1)},a^{(j+1)}; \theta)$ \;
        Perform Stochastic Gradient Descent update on $\theta$ according to the loss function (equation \ref{Weighted-mean-square-loss}) \;
        }
        
  \caption{Decision\_Maker\_Agent}
  \label{alg:Decider-Agent-DQN-Algorithm}
\end{algorithm}

\section{Experiments And Evaluation Results} \label{sec:exp}

In this section, we will evaluate the effectiveness of the proposed IDS by conducting a comprehensive assessment on the CIC-IDS-2017 \cite{sharafaldin_toward_2018} dataset. Firstly, we will describe the preprocessing steps that have been applied to the dataset. Subsequently, we will present the evaluation results of our IDS, considering various metrics. Finally, we will provide a comparative analysis with state-of-the-art works to gain further insights into the performance of our system.


\subsection{Dataset Preprocessing}
The CIC-IDS-2017 \cite{sharafaldin_toward_2018} dataset is a labeled network traffic dataset that includes advanced contemporary attack techniques.  As shown in Table \ref{tab:CIC-IDS-2017}, the dataset exhibits a highly imbalanced class distribution, with the majority of instances classified as BENIGN, whereas certain attack classes have a significantly limited number of instances.
To preprocess the dataset, we first delete records with infinite or missing values to ensure stability during training. Next, we apply Z-score normalization to scale the features, ensuring they exhibit a mean of zero and a standard deviation of one. It is a common technique used for standardizing data.  Finally, as shown in Table \ref{tab:CIC-IDS-2017} we split the dataset into a training set and a testing set, with 80\% of the data used for training and 20\% used for testing. We have also decreased the number of normal traffic samples in order to balance it with the number of malicious traffic samples.

\begin{table}[h]
  \centering
  \caption{CIC-IDS-2017 dataset}
  \label{tab:CIC-IDS-2017}
\begin{adjustbox}{max width=\columnwidth}
  \begin{tabular}{|l|c|c|c|c|}
    \hline
    \textbf{Class} & \textbf{Total} & \textbf{Preprocessed} & \textbf{Training} & \textbf{Testing}\\
    \hline
    BENIGN & 2271781 & 2271320 & 559999 & 140001 \\
    \hline
    DoS Hulk & 230124 & 230124 & 184099 & 46025 \\
    \hline
    PortScan & 158804 & 158804 & 127043 & 31761 \\
    \hline
    DDoS & 128027 & 128025 & 102420 & 25605 \\
    \hline
    DoS GoldenEye & 10293 & 10293 & 8234 & 2059 \\
    \hline
    FTP Patator & 7938 & 7935 & 6348 & 1587 \\
    \hline
    SSH Patator & 5897 & 5897 & 4717 & 1180 \\
    \hline
    DoS slowloris & 5796 & 5796 & 4637 & 1159 \\
    \hline
    DoS Slowhttptest & 5499 & 5499 & 4399 & 1100 \\
    \hline
    Bot & 1956 & 1956 & 1565 & 391 \\
    \hline
    Web\_Attack Brute Force & 1507 & 1507 & 1206 & 301 \\
    \hline
    Web\_Attack XSS & 652 & 652 & 522 & 130 \\
    \hline
    Infiltration & 36 & 36 & 29 & 7 \\
    \hline
    Web\_Attack Sql Injection & 21 & 21 & 17 & 4 \\
    \hline
    Heartbleed & 11 & 11 & 9 & 2 \\
    \hline
  \end{tabular}
\end{adjustbox}
\end{table}

\subsection{Performance evaluation of the proposed model}
To assess the effectiveness of our proposed multi-agent RL-based detection model, we consider the following metrics:

\vspace{8pt}
$\text{Accuracy} = \frac{TP + TN}{TP + TN + FP + FN}$  $\text{Precision} = \frac{TP}{TP + FP}$

\vspace{2pt}

$\text{Recall} = \frac{TP}{TP + FN}$ $\text{F1-Score} = 2 \cdot \frac{Precision \cdot Recall}{Precision + Recall}$

\vspace{2pt}

$\text{False Positive Rate (FPR)} = \frac{FP}{FP + TN}$
\vspace{8pt}

\noindent TP, TN, FP, and FN denote true positive, true negative, false
positive, and false negative, respectively.


The hyperparameters for both the $L_1$\_agents and the decision-maker\_agent are provided in Table \ref{tab:Hyperparameters}. We adopted the same architecture across all agents, including the decision-maker, to ensure simplicity and efficiency.
\begin{table}[h!]
    \centering
    \caption{DQN hyperparameters values}
    \begin{tabular}{|c|c|}
        \hline        
         Replay buffer size  & 10000000 \\
        \hline
         Minibatch size & 1000000 \\
         \hline
         Activation functions & [ReLU, ReLU, Linear]  \\
        \hline
         Optimizer & Adam \\
        \hline
         Learning rate & 0.01 \\
        \hline
         Gamma for Q-Values & 0.01 \\
        \hline
         Epsilon decay & 0.01 \\
        \hline 
         Hidden layers & 2 \\ 
        \hline
         No. of Neurons & [128, 128] \\ 
        \hline
         episodes & 300 \\ 
        \hline
    \end{tabular}
    \label{tab:Hyperparameters}
\end{table}

We tested our approach on the 14 classes of attacks, in contrast to the majority of research articles, which group together attack classes of the same category, e.g. in the web attack category there are 3 classes, and the DoS category includes several sub-classes also. This was done in order to test the ability of our model to detect patterns and differences, even on classes that share similarities and have limited data samples. 
Based on the testing set results presented in Table \ref{tab:performance}, the proposed IDS demonstrated exceptional detection capabilities ( 99\% accuracy). It achieved a high detection rate ( 99\% weighted recall) and a low false positive rate (0.0016 \%). The proposed IDS effectively detects a wide range of attacks, including those with a limited number of samples. This demonstrates the model's ability to address the class imbalance problem in the CIC-IDS-2017 dataset and provide precise classification of attacks. Not only does it accurately detect each class of attack, but it also maintains high performance in identifying benign traffic.


\begin{table}[h]
\centering
\caption{Performance Measures on CIC-IDS-2017 Dataset}
\begin{adjustbox}{max width=\columnwidth}
\begin{tabular}{lccccc}
\hline
\textbf{Classes} & \textbf{Precision} & \textbf{Recall} & \textbf{F1-Score} & \textbf{FPR} & \textbf{Support} \\
\hline
BENIGN & 1 & 0.99 & 0.99 & 0.00145 & 140001\\
DoS Hulk & 0.99 & 1 & 1 & 0.00102 & 46025 \\
PortScan & 0.99 & 1 & 1 & 0.00009 & 31761\\
DDoS & 1 & 1 & 1 & 0.00007 & 25605\\
DoS GoldenEye & 0.99 & 0.99 & 0.99 & 0.00002 & 2059\\
FTP Patator & 1 & 0.98 & 0.99 & 0.00008 & 1587\\
SSH Patator & 0.98 & 0.97 & 0.98 & 0.00003 & 1180 \\
DoS slowloris & 0.99 & 0.99 & 0.99 & 0.00017 & 1159 \\
DoS Slowhttptest & 0.96 & 0.99 & 0.98 & 0.00270 & 1100 \\
Bot & 0.35 & 0.92 & 0.51 & 0.00029 & 391 \\
Web Attack Brute Force & 0.73 & 0.64 & 0.68 & 0.00046 & 301 \\
Web Attack XSS & 0.4 & 0.59 & 0.48 & 0.00002 & 130 \\
Infiltration & 0.33 & 0.43 & 0.38 & 0.00056 & 7\\
Web Attack Sql Injection & 0.01 & 0.5 & 0.03 & 0.00001 & 4 \\
Heartbleed & 0.33 & 0.5 & 0.4 & 0.00211 & 2 \\
\textbf{Weighted Average} & \textbf{0.99} & \textbf{0.99} & \textbf{0.99} & \textbf{0.0016} & \textbf{251312} \\
\hline
\textbf{Accuracy} & \textbf{0.99} \\
\textbf{AUC} & \textbf{0.92}  \\
\hline

\end{tabular}
\end{adjustbox}
\label{tab:performance}
\end{table}

\begin{figure}[hbt!]
  \centering
  \includegraphics[width=9cm]{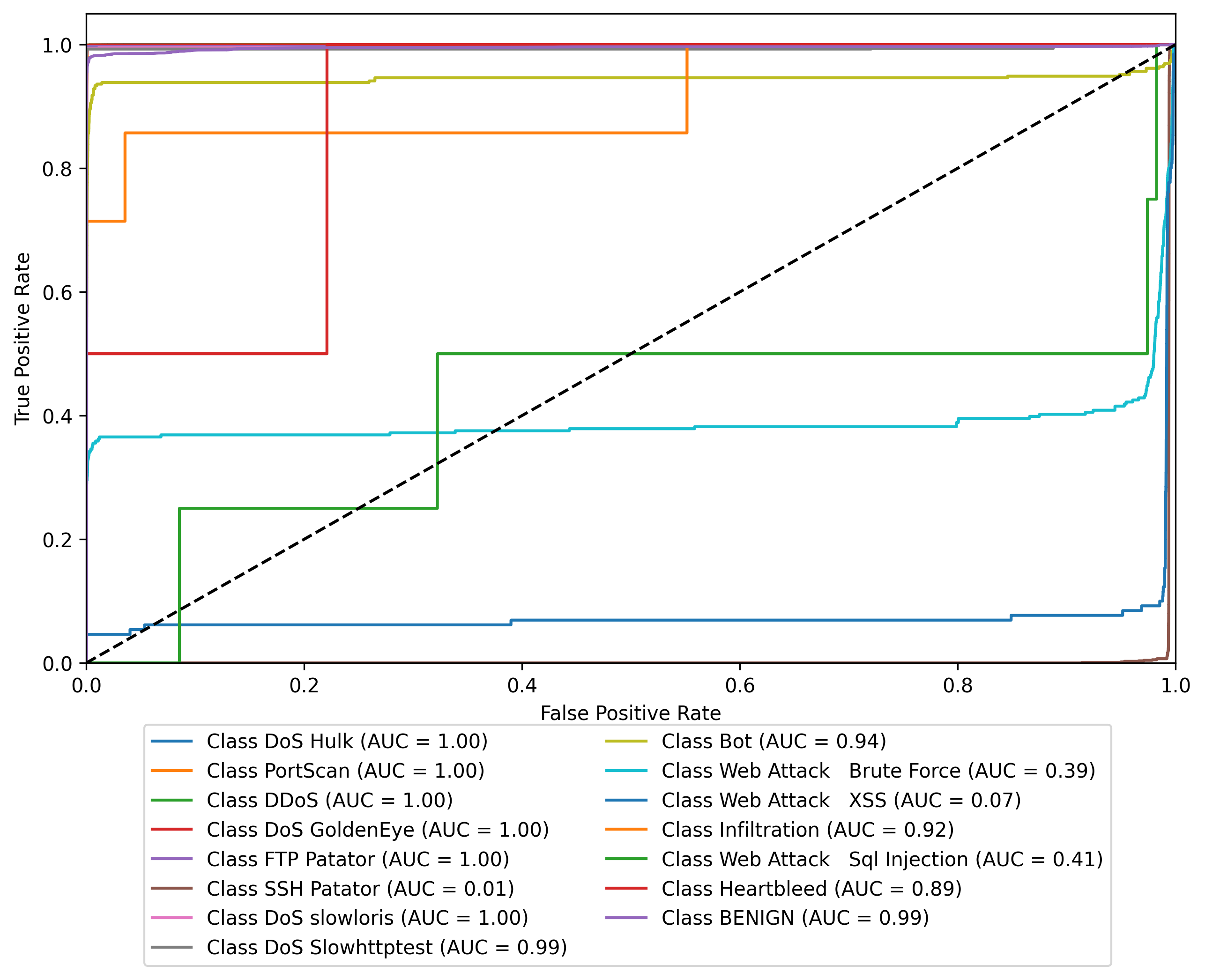}
  \caption{ROC Curve - CIC-IDS-2017}
  \label{fig:ROC-CURVE}
\end{figure}

The ROC curve and AUC values for each attack type in the CIC-IDS-2017 dataset are depicted in Figure \ref{fig:ROC-CURVE}. Our IDS demonstrates excellent AUC results for the majority of the classes, with the exception of four classes out of the total fifteen. These four classes consist of three web attacks, which have a notably smaller number of samples compared to the other classes. Furthermore, for the SSH Patator class, the reduced AUC value can be attributed to the low data quality of this particular class within the CIC-IDS-2017 dataset. Consequently, the model may encounter challenges in learning representative patterns, leading to a decrease in the AUC value.
The global AUC value of $92\%$ is given also in the Table \ref{tab:performance}. These results confirm that our model is capable of distinguishing between classes and provides a fine-grained classification.

\begin{figure}[hbt!]
  \centering
  \includegraphics[width=9cm]{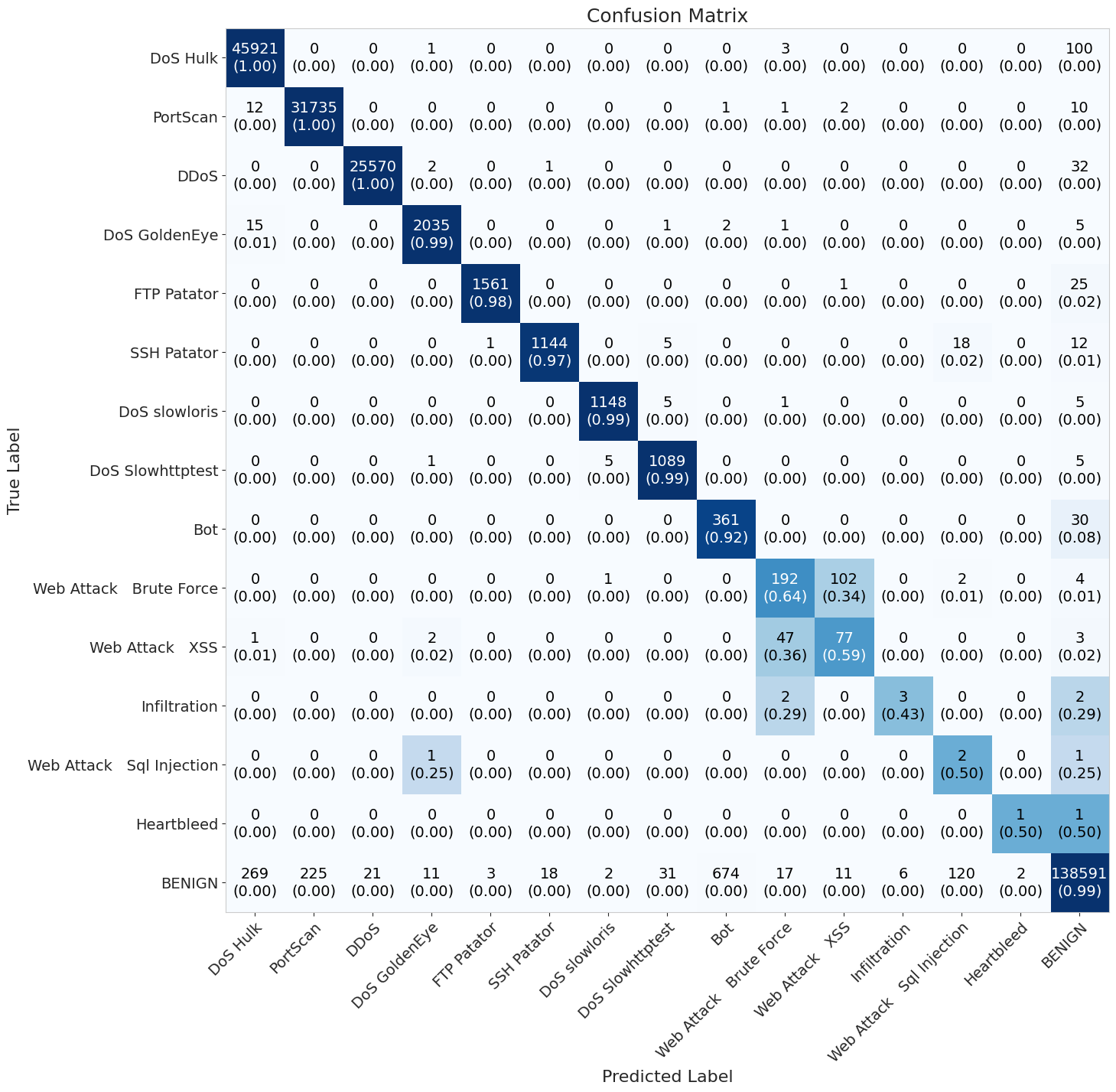}
  \caption{Confusion matrix}
  \label{fig:Confusion-matrix}
\end{figure}

The confusion matrix depicted in Figure \ref{fig:Confusion-matrix} illustrates the performance of our IDS. The true labels are represented on the Y-axis, while the predicted labels are displayed on the X-axis. Overall, our IDS demonstrates accurate classification of majority of attack types. However, some attack types, such as Web attacks, Infiltration, and Heartbleed, present a challenge due to their significantly smaller sample sizes compared to the other attack types. Despite these challenges, our IDS still achieves acceptable results, with accuracy rates surpassing 50\% for these classes. For instance, in the case of the Heartbleed attack, our testing dataset only contains two samples, yet the IDS successfully detected one sample of this attack. Similarly, when it comes to web attacks, the majority of false negatives are misclassified within another web attack type. This implies that our IDS is capable of distinguishing web attacks from other classes; however, it struggles to differentiate between different types of web attacks due to the limited amount of data available for these specific attack types.


We have conducted experiments to validate the evolvability and robustness of our architecture when faced with new attacks and its ability to effectively adapt to changes in existing attack patterns. To achieve this, we categorized the 15 attack classes into 7 groups: (D)DOS (including DoS Hulk, DDOS, DoS GoldenEye, DoS slowloris, and DoS Slowhttptest), Port Scan, Brute Force (FTP-Patator, SSH-Patator), Web Attacks (comprising Web Attack – Brute Force, Web Attack – XSS, Web Attack – SQL Injection) , Infiltration, and Heartbleed. In order to simulate the evolution of attack patterns, we trained the system by initially excluding a sub-class of DOS, specifically DoS slowloris, and a class of web attack, Web Attack – XSS, from the training set while keeping them in the testing set. We note that the choice of excluded attacks is totally random Subsequently, we conducted a second training adaptation where we exclusively trained two agents from the first level, specifically the DOS and web attacks agents, along with the decider agent. This training involved the use of the previous training dataset combined with $80\%$ of the newly acquired data on a small span of 20 episodes. The remaining $20\%$ of data was used for testing. 

\begin{table*}[h]
\centering
\caption{Performance Measures for Evaluating the Evolvability of the Proposed Model on the CIC-IDS-2017 Dataset}
\begin{adjustbox}{max width=\textwidth}
\begin{tabular}{lccccccccc}
\hline
 & \textbf{Before adaptation} &  &  &  & \textbf{After adaptation} & & & & \\

\hline
\textbf{Classes} & \textbf{Precision} & \textbf{Recall} & \textbf{F1-Score} & \textbf{FPR} & \textbf{Precision} & \textbf{Recall} & \textbf{F1-Score} & \textbf{FPR} & \textbf{Support} \\
\hline
BENIGN & 0.99 & 0.99 & 0.99 & 0.01 & 1.00 & 0.99 & 0.99 & 0.003 & 140001\\
(D)DoS & 0.99 & 0.99 & 0.99 & 0.003 & 0.99 & 0.99 & 0.99 & 0.004 & 75948\\
PortScan & 0.99 & 1.00 & 1.00 & 0.001 & 0.99 & 1.00 & 0.99 & 0.001 & 31761\\
Brute Force & 0.95 & 0.92 & 0.94 & 0.0005 & 0.95 & 0.97 & 0.96 & 0.0005 & 2767\\
Bot & 0.23 & 0.85 & 0.37 & 0.004 & 0.33 & 0.91 & 0.48 & 0.002 & 391\\
Web Attacks & 0.98 & 0.91 & 0.95 & 0.00003 & 0.98 & 0.88 & 0.92 & 0.00003 & 435\\
Infiltration & 0.29 & 0.71 & 0.42 & 0.00005 & 0.19 & 0.71 & 0.30 & 0.00008 & 7\\
Heartbleed & 0.25 & 0.50 & 0.33 & 0.00001 & 0.5 & 0.5 & 0.5 & 0.000004 & 2\\
\textbf{Weighted Average} & \textbf{0.99} & \textbf{0.99} & \textbf{0.99} & \textbf{0.0071} & \textbf{0.99} & \textbf{0.99} & \textbf{0.99} & \textbf{0.0035} & \textbf{251312}\\
\hline
DoS slowloris & - & 0.33 & - & - & - & 0.98 & - & - & 1159\\
Web Attack XSS & - & 0.91 & - & - & - & 0.92 & - & - & 130\\
\hline
\textbf{Accuracy Before adaptation} & \textbf{0.99} &  &  & & \textbf{Accuracy After adaptation} & \textbf{0.99}  &  &  & \\
\hline
\end{tabular}
\end{adjustbox}
\label{tab:Adaptation}
\end{table*}

The testing set results of the evolvability case study are presented in Table \ref{tab:Adaptation}. Based on the performance measures of our model before and after adaptation, significant improvements can be observed in the detection rate (recall) of the two excluded classes, namely DoS Slowloris and Web Attack – XSS.
For DoS slowloris, the recall (detection rate) has notably increased from 0.33 to 0.98 after adaptation. Similarly, for Web Attack – XSS, the recall has improved marginally from 0.91 to 0.92. For Web Attack – XSS, we can observe that the detection rate before adaptation has been quite high because it closely resembles other web attacks.
It is important to highlight that the overall performance of our model has not deteriorated as a result of adaptation. 
Furthermore, we observed improvements in some classes and a slight degradation in others. We also noted a decrease in the overall false positive rate.
\subsection{Performances comparison with the state-of-the-art works}
\begin{figure}[hbt!]
  \centering
  \includegraphics[width=9cm]{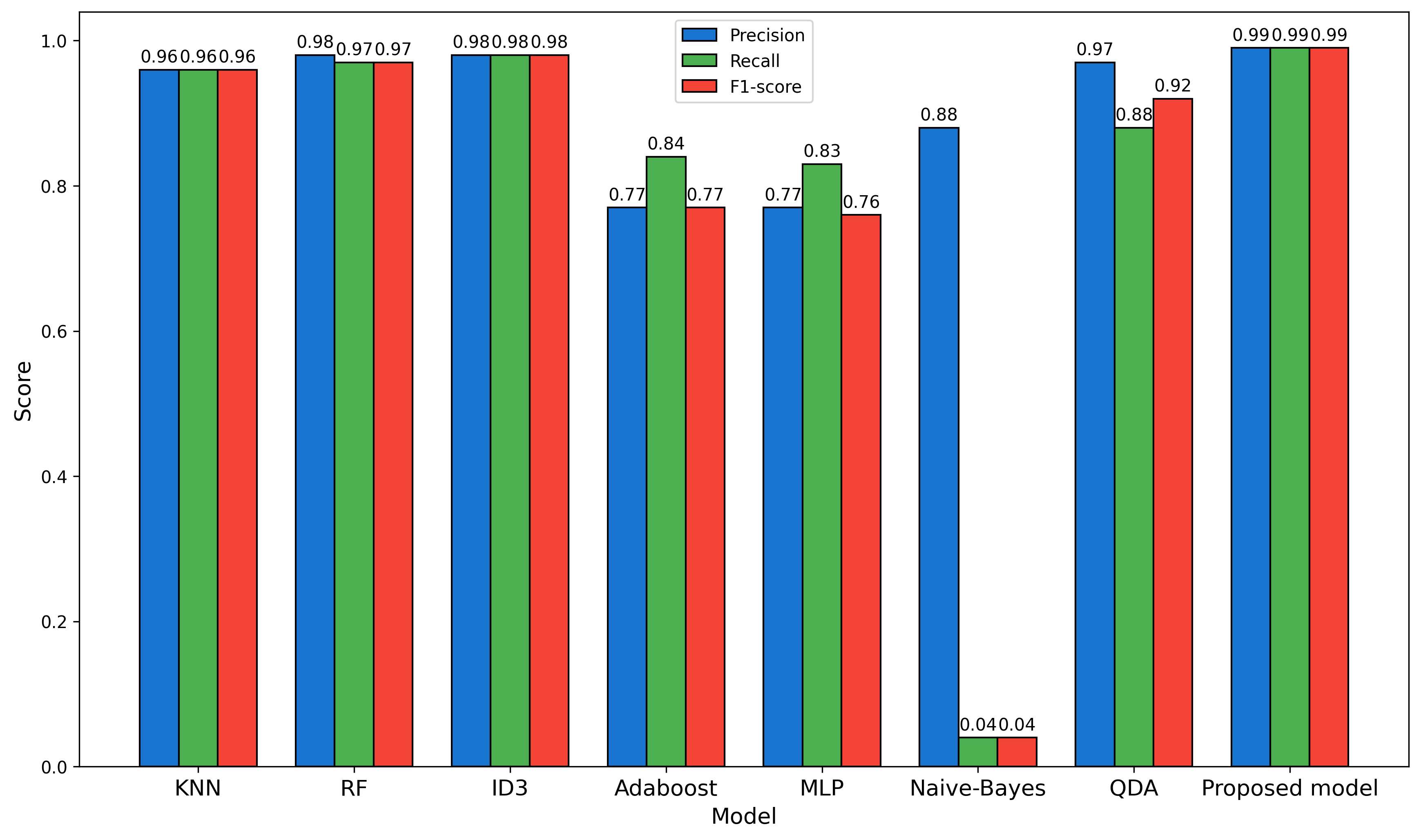}
  \caption{Comparison of our proposed model with other machine learning models}
  \label{fig:bar-chart-proposed-model-vs-ML}
\end{figure}

We performed a comprehensive performance analysis, comparing our proposed IDS with existing state-of-the-art solutions that leverage the CIC-IDS-2017 dataset. In a study conducted by Sharafaldin et al. \cite{sharafaldin_toward_2018}, various machine learning algorithms were evaluated on the CIC-IDS-2017 dataset. These algorithms included K-Nearest Neighbors (KNN), Random Forest (RF), ID3, Adaboost, Multilayer Perceptron (MLP), Naive-Bayes (NB), and Quadratic Discriminant Analysis (QDA). Figure \ref{fig:bar-chart-proposed-model-vs-ML} depicts the evaluation of the proposed IDS in comparison to the previously mentioned ML algorithms. The assessment is based on three key performance metrics: precision, recall, and F1 score. Our IDS demonstrates superior performance across all metrics, surpassing all other machine learning algorithms.


We further conducted a comprehensive evaluation of our proposed IDS by comparing it with state-of-the-art IDS models \cite{ramana_ambient_2022, sethi_attention_2021} that employ reinforcement learning techniques and utilize the CIC-IDS-2017 dataset. The comparative analysis, as shown in Table \ref{tab:performance2}, clearly demonstrates that our IDS outperforms the existing approaches, achieving remarkable results. Furthermore, our solution demonstrates an exceptional capability to effectively detect all attacks including minor classes, as well as accurately identifying benign traffic, resulting in an impressively low false positive rate of 0.0016 compared to \cite{sethi_attention_2021}.


\begin{table}[H]
\centering
\caption{Comparison of performance: Our model vs. State-of-the-art works on CIC-IDS-2017}
\begin{adjustbox}{max width=\columnwidth}
\begin{tabular}{lccccc}
\hline
\textbf{Reference} & \textbf{Accuracy} & \textbf{Precision} & \textbf{Recall} & \textbf{F1-Score} & \textbf{FPR ($\%$)} \\
\hline
RL-DQN \cite{ramana_ambient_2022} & 0.97 & - & - & - & -\\
\hline
A-DQN \cite{sethi_attention_2021} & 0.98 & 0.98 & 0.99 & 0.98 & 0.82 \\
\hline
Proposed Model & \textbf{0.99} & \textbf{0.99} & \textbf{0.99} & \textbf{0.99} & \textbf{0.16}\\
\hline
\end{tabular}
\end{adjustbox}
\label{tab:performance2}
\end{table}

\section{Conclusion} \label{sec:con}
This paper introduced a novel network intrusion detection system. The proposed system leveraged an innovative multi-agent deep reinforcement learning architecture, comprising two levels of reinforcement learning. The initial detection level comprises N independent RL agents, each specializing in detecting a specific type of attack. These agents are then followed by a decision-maker agent that provides the final classification. The proposed solution features a flexible and evolvable architecture, enabling the addition of new attacks and the effective handling of evolving attack patterns. To enhance the Deep Q network algorithm, we employ the weighted mean square loss function and adopt cost-sensitive learning techniques to address class imbalance issues. Our performance evaluation demonstrates that our model effectively tackles the class imbalance problem and achieves a fine-grained classification of attacks, yielding an impressively low false positive rate.

Several mechanisms have emerged to enable the sharing of Cyber Threat Intelligence (CTI) information to address large-scale attacks. As future lines of work, we plan to extend the present solution into a decentralized machine learning model to ensure collaboration in the intrusion detection process.

\bibliographystyle{IEEEtran}
\bibliography{IEEEabrv,Bibliography}

\vspace{12pt}

\end{document}